\documentclass[cits]{PoS}

\usepackage{amsmath,amssymb,amsthm,amsfonts}
\usepackage{exscale}
\usepackage{mathtools}

\usepackage{bbm}
\usepackage{lmodern}
\usepackage{shuffle}

\usepackage{graphicx}

\usepackage{tabularx}
\usepackage{booktabs}

\newcommand{\N}{
\mathbbm{N}
}
\newcommand{\Z}{
\mathbbm{Z}
}
\newcommand{\R}{
\mathbbm{R}
}
\newcommand{\Rplus}{\R_{+}}
\newcommand{\C}{
\mathbbm{C}
}

\newlength{\wurelwidth}

\newcommand{\wurel}[2][=]{\mathrel{\mathop{#1}_{\!\scalebox{0.5}{\makebox[\the\wurelwidth]{#2}}\!}}}

\DeclareMathOperator{\sdd}{\omega}

\newcommand{\dimension}{D}
\newcommand{\Graph}[2][1.0]{%
\vcenter{\hbox{\includegraphics[scale=#1]{Graphs_#2}}}%
}

\newcommand{\phipol}{\varphi}
\newcommand{\psipol}{\psi}

\newcommand{\hide}[1]{}

\newcommand{\dd}[1][]{\mathrm{d}^{#1}}
\newcommand{\restrict}[2]{%
{\left. #1 \right|}_{#2}%
}

\newcommand{\defas}{
\mathrel{\mathop:}=
}

\newcommand{\set}[1]{
\left\{ #1 \right\}
}

\newcommand{\SP}{\alpha}
\newcommand{\EPE}{\varepsilon}
\newcommand{\EP}{\nu}

\DeclareMathOperator{\Li}{Li}
\newcommand{\cupdot}{\mathbin{\dot{\cup}}}
\newcommand{\mzv}[2][]{\zeta_{#2}^{#1}}

\newcommand{\physical}[1]{\widehat{#1}}

\newcommand{\zigzag}[1]{\text{\upshape ZZ}_{#1}}

\newcommand{\Maple}{{\ttfamily Maple}}

\title{Feynman integrals via hyperlogarithms}

\ShortTitle{Feynman integrals via hyperlogarithms}

\author{\speaker{Erik Panzer}%
\\%
	Institues of Physics and Mathematics,
	Humboldt-Universit\"{a}t zu Berlin\\
	Unter den Linden 6, 10099 Berlin, Germany\\%
	E-mail: \email{panzer@mathematik.hu-berlin.de}%
}

\abstract{%
This talk summarizes recent developments in the evaluation of Feynman integrals using hyperlogarithms. We discuss extensions of the original method, new results that were obtained with this approach and point out current problems and future directions.
}

\FullConference{Loops and Legs in Quantum Field Theory\\
                 27 April 2014 -- 02 May 2014\\
                 Weimar, Germany}

\begin{document}

\section{Schwinger parameters and hyperlogarithms}
Many talks at this conference demonstrate the remarkable progress in the exact evaluation of Feynman integrals that was achieved during the past few years. A key element shared by many of these advances is an improved understanding of multiple polylogarithms
\begin{equation}
	\Li_{n_1,\ldots,n_r}(z_1,\ldots,z_r)
	= \sum_{0<k_1<\ldots< k_r} \frac{z_1^{k_1}\cdots z_r^{k_r}}{k_1^{n_1}\cdots k_r^{n_r}},
	\quad\text{where}\quad
	n_1,\ldots,n_r \in \N,
	\label{eq:def-Li} %
\end{equation}
which suffice to express a wide class of Feynman integrals.\footnote{Counterexamples are known in massive \cite{AdamsBognerWeinzierl:Sunrise,BlochVanhove:Sunset}, massless \cite{BrownSchnetz:ZigZag} and even supersymmetric theories \cite{CaronHuotLarsen:UniquenessTwoLoopMasterContours,NandanPaulosSpradlinVolovich:StarIntegrals}.} They admit representations as iterated integrals \cite{Chen:IteratedPathIntegrals} of the form
\begin{equation}
	L_{\sigma_n,\ldots,\sigma_1}(z)
	= \idotsint\limits_{0 < z_1 < \ldots z_n<z} \frac{\dd z_n}{z_n - \sigma_n} \cdots \frac{\dd z_1}{z_1 - \sigma_1}
	\quad\text{with}\quad
	\sigma_1,\ldots,\sigma_n \in \C
	\quad\text{and}\quad
	\sigma_1 \neq 0,
	\label{eq:def-hyper} %
\end{equation}
called hyperlogarithms \cite{LappoDanilevsky}, and can be manipulated symbolically very efficiently.
In these proceedings we consider the method put forward by Francis Brown \cite{Brown:TwoPoint}, that aims at integrating out Schwinger parameters $\SP_e$ one by one in the representation
\begin{equation}
	\Phi(G)
	= \frac{\Gamma(\sdd)}{\prod_e \Gamma(\EP_e)}
		\int_{\Rplus^{N}}
		\frac{\delta(1-\SP_{e_N})}{\psipol^{\dimension/2}} \left( \frac{\psipol}{\phipol} \right)^{\sdd}
		\prod_{e \in E} \SP_e^{\EP_e-1}\dd \SP_e,
	\quad
	\sdd 
	= \sum_{e \in E} \EP_e - \frac{\dimension}{2} h_1(G) 
	\label{eq:projective-rep} %
\end{equation}
of Feynman integrals associated to a graph $G$ with $h_1(G)$ loops and $N$ edges $E$. This formula assumes scalar propagators $(k_e^2 + m_e^2)^{-\EP_e}$ of mass $m_e$ at each edge $e$ (raised to some power $\EP_e$ called \emph{index}), but generalizations to tensor integrals exist \cite{Nakanishi:GraphTheoryFeynmanIntegrals}.
The \emph{Symanzik polynomials} \cite{BognerWeinzierl:GraphPolynomials} sum all spanning trees $T$ and 2-forests $F$ in
\begin{equation}
	\psipol = \sum_T \prod_{e \notin T} \SP_e
	\quad\text{and}\quad
	\phipol = \sum_F q(F)^2 \prod_{e \notin F} \SP_e + \psipol \sum_{e \in E} \SP_e^{} m_e^2
	,
	\label{eq:def:polynomials} %
\end{equation}
where $q(F)^2$ is the square of the momentum flowing between the two components of $F$.
We choose an order $e_1,\ldots,e_N$ of edges and compute
$ I_k \defas \int_0^{\infty} I_{k-1}\ \dd \SP_{e_k}$ iteratively,
starting with the original integrand $I_0$ of \eqref{eq:projective-rep}. Details of the involved algorithms are given in \cite{Brown:PeriodsFeynmanIntegrals,Brown:TwoPoint} and a public implementation in {\Maple} \cite{Maple} is available \cite{Panzer:HyperIntAlgorithms}. Also the more general approach presented at the preceeding conference \cite{BognerBrown:SymbolicIntegration}, based on iterated integrals in several variables (instead of just one), is going to be published in the nearest future.

The crucial limitation of this method is that all $I_k(\SP_{e_{k+1}}, \ldots, \SP_{e_N})$ must be expressible as $\C(\SP_{e_{k+1}},\ldots,\SP_{e_N})$-linear combinations hyperlogarithms \eqref{eq:def-hyper} with respect to the next integration variable $z = \SP_{e_{k+1}}$. If this holds for some order on $E$, we call $G$ \emph{linearly reducible}. This criterion depends only on the Symanzik polynomials and reduction algorithms are available \cite{Brown:PeriodsFeynmanIntegrals,Brown:TwoPoint} that check conditions sufficient for linear reducibility.\footnote{This means that reductions can reveal linear reducibility, but not disprove it.}

In particular this discussion is unaffected by infinitessimal expansions in analytic regulators, like the popular shift $\dimension = \physical{\dimension} - 2\varepsilon$ away from an even dimension $\physical{\dimension} \in 2\N$ of spacetime%
\footnote{Even if one starts out with $\physical{\dimension} = 4$ dimensions, a reduction of tensor integrals will produce scalar integrals also in even dimensions above $\physical{\dimension}$.}
such that
$
	\Phi(G)
	=
	\sum_{k} \Phi_{k}(G) \cdot \varepsilon^k
$
becomes a Laurent series. All of its coefficients can in principle be computed as soon as $G$ is linearly reducible, and remarkably this also applies for the multivariate expansions in the indices
$\EP_e = \physical{\EP}_e + \EPE_e$ close to integers $\physical{\EP}_e \in \Z$.

\section{Massless propagators up to $6$ loops}
\begin{figure}\centering%
	\begin{tabular}{cccccc}
		$\Graph[0.25]{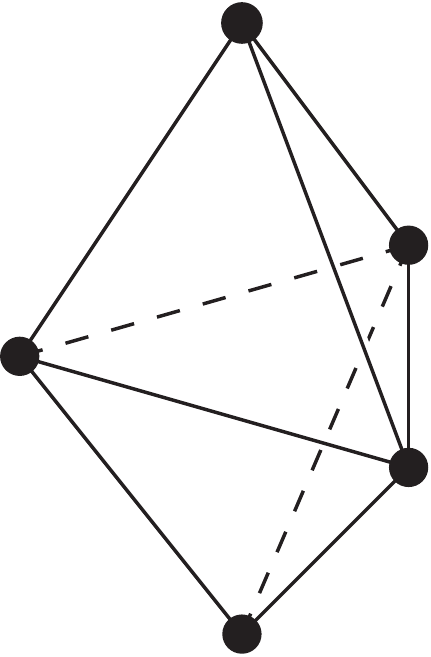}$ &
		$\Graph[0.3]{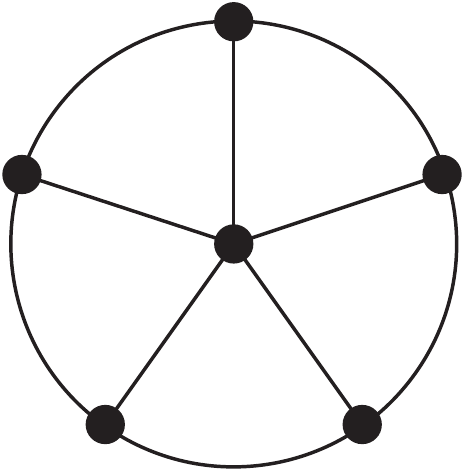}$ &
		$\Graph[0.4]{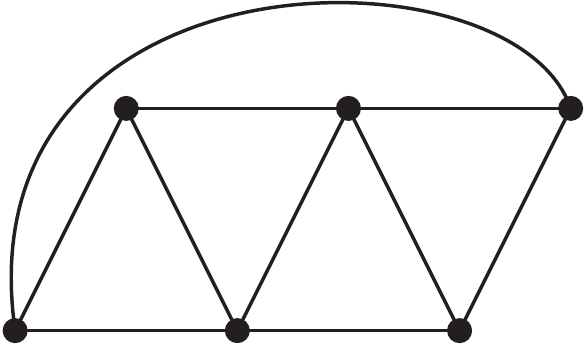}$ &
		$\Graph[0.4]{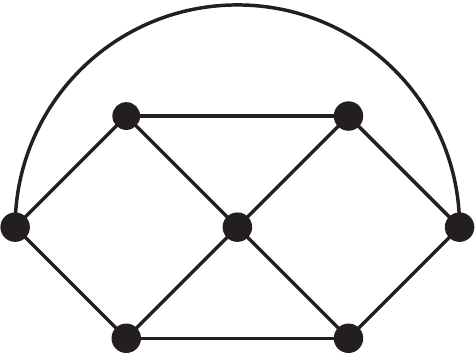}$ &
		$\Graph[0.4]{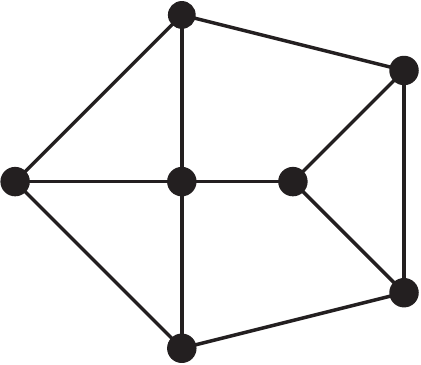}$ &
		$\Graph[0.4]{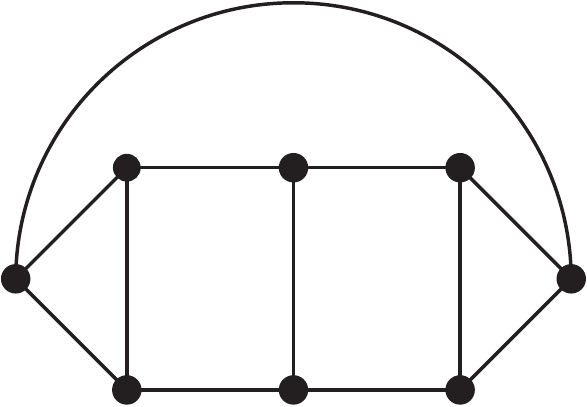}$ \\
		${_5 P_1}$ &  ${_5 P_2}$ & ${_5 P_3}$ & ${_5 P_4}$ & $ {_5 P_5}$ & ${_5 P_6}$\\
	\end{tabular}\\%
	\begin{tabular}{ccccc}
		$\Graph[0.37]{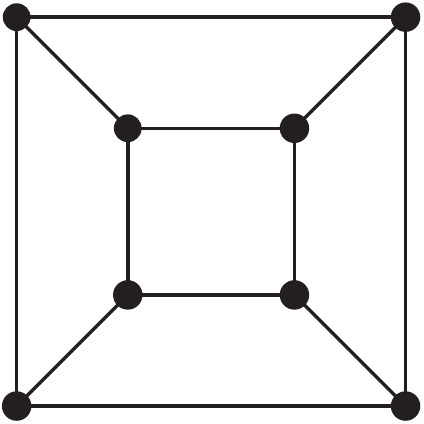}$ &
		$\Graph[0.4]{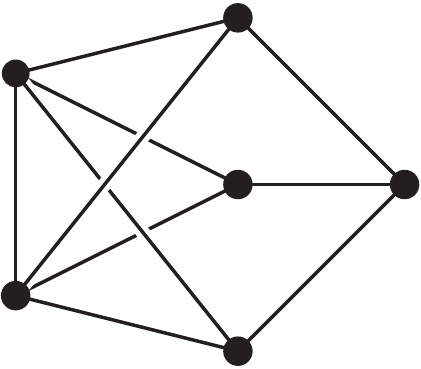}$ &
		$\Graph[0.4]{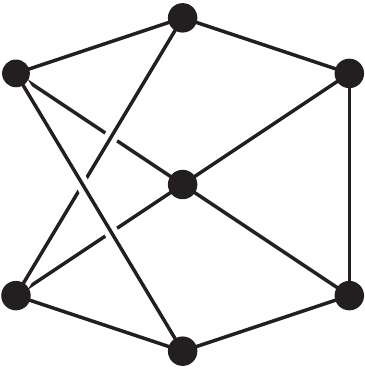}$ &
		$\Graph[0.35]{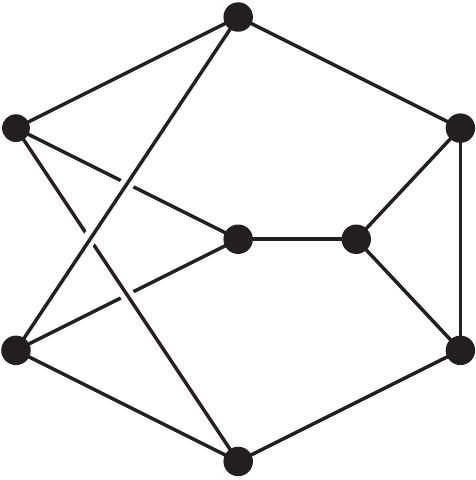}$ &
		$\Graph[0.38]{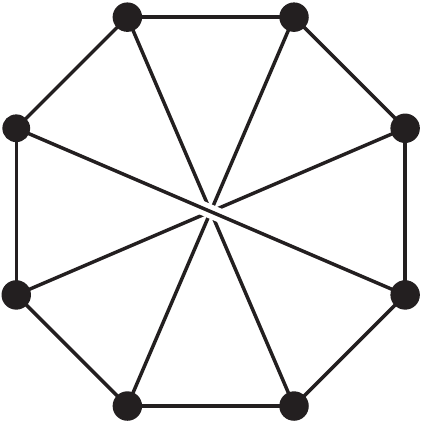}$\\
		${_5 P_7}$ & ${_5 N_1}$ &$ {_5 N_2} $ & ${_5 N_3}$ & ${_5 N_4}$ \\
	\end{tabular}%
	\caption{%
	All three-connected five-loop vacuum graphs \cite{Panzer:MasslessPropagators}, divided into planar ($P$) and non-planar ($N$) ones. 
	The zig-zag ${_5 P_3} = \zigzag{5}$ and ${_5 N_1}$ were considered in \cite{Brown:TwoPoint}. 
	Cutting any edge produces a propagator with four loops, deleting a three-valent vertex creates a three-loop three-point graph.}%
	\label{fig:5loop-vacuum} %
\end{figure}%
\begin{figure}%
	\begin{gather*}%
		P_{7,8}
		=
		\Graph[0.34]{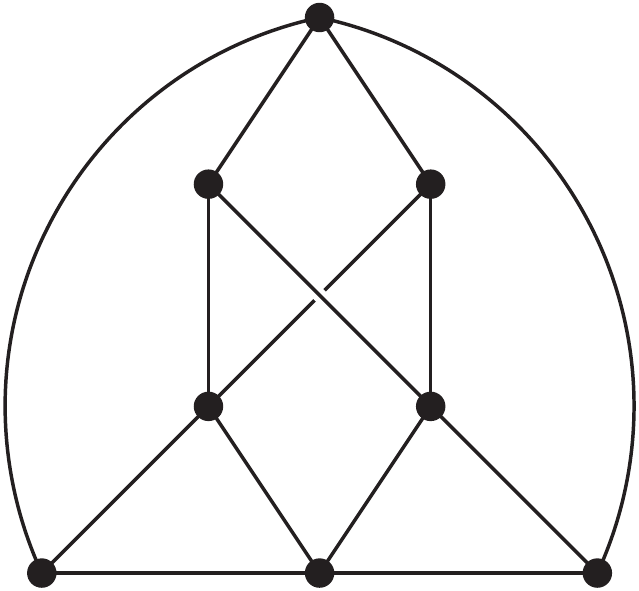}
		\qquad
		P_{7,9}
		=
		\Graph[0.4]{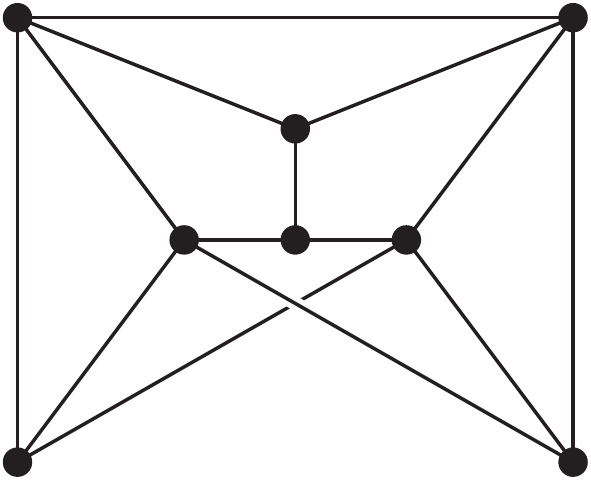}
		\qquad
		P_{7,11}
		=
		\Graph[0.4]{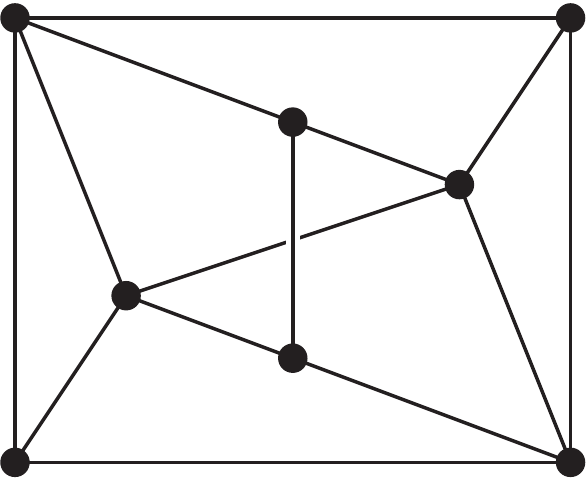}
	\end{gather*}\vspace{-7mm}%
	\caption{The most complicated primitive periods in $\phi^4$-theory at seven loops are given by the three graphs $P_{7,8}$, $P_{7,9}$ and $P_{7,11}$ in the notation of the census \cite{Schnetz:Census}.}%
	\label{fig:7loop-primitives}%
\end{figure}%
Originally, this method was applied to convergent massless propagators $G$ only \cite{Brown:TwoPoint}. These transform into vacuum graphs $\widetilde{G} = G \cupdot \set{e}$ upon joining the external legs of $G$ by an edge $e$. Then only the first Symanzik $\psipol_{\widetilde{G}} = \psipol_G^{} \SP_e + \phipol_G^{}$ is of interest and a powerful toolbox of algebraic identities becomes available \cite{Brown:PeriodsFeynmanIntegrals,BrownSchnetz:ZigZag,BrownYeats:SpanningForestPolynomials}.

We verified that all massless propagators up to four loops (e.\,g.\ all which arise by cutting an edge of the graphs shown in figure~\ref{fig:5loop-vacuum}) are linearly reducible \cite{Panzer:MasslessPropagators}. The recent computation \cite{BaikovChetyrkin:FourLoopPropagatorsAlgebraic,SmirnovTentyukov:FIESTA} of the corresponding master integrals (using different methods) can therefore be extended to higher orders in $\varepsilon$ and to include self-energy insertions. Such results will be needed for calculations at higher loop orders and are already available for many examples \cite{Panzer:MasslessPropagators}.

The traditional benchmark in this field is the evaluation of primitive divergent periods of $\phi^4$-theory in $\dimension=4$ dimensions \cite{BroadhurstKreimer:KnotsNumbers}. This problem is now completely solved up to 7 loops: All primitive graphs of the census \cite{Schnetz:Census} to this order are linearly reducible---except for $P_{6,4}$ (that can be computed with \emph{graphical functions} \cite{Schnetz:GraphicalFunctions}) and $P_{7,11}$ of figure~\ref{fig:7loop-primitives} which we address below. These results suggest that all massless propagators with at most $6$ loops could be computable using multiple polylogarithms. This is the most optimistic scenario, as some $7$ loop massless propagators are known to exceed the world of polylogarithms \cite{BrownSchnetz:K3phi4}.

\subsection{Changing variables for linear reductions}
After ten integrations for $P_{7,11}$, the partial integral $I_{10} = L / d_{10}$ consists of a hyperlogarithm $L$ of weight $8$ and the irreducible, totally quadratic denominator
\begin{equation}\begin{split}
d_{10} &=
\SP_2 \SP_4^2 \SP_1 + \SP_2 \SP_4^2 \SP_3 - \SP_1 \SP_2 \SP_3\SP_4 + \SP_2^2 \SP_4\SP_1 + \SP_2^2 \SP_4\SP_3
- 2 \SP_2\SP_3^2\SP_4 - \SP_2^2\SP_3^2
\\ &\quad
- 2\SP_2^2\SP_3\SP_1 - 2\SP_2\SP_3^2\SP_1 - \SP_3^2\SP_4^2 
- 2\SP_3^2\SP_4\SP_1 - \SP_2^2\SP_1^2 - 2\SP_2\SP_3\SP_1^2 - \SP_3^2\SP_1^2
.
	\label{eq:P711-d10}%
\end{split}\end{equation}
Further integration would introduce square roots of the discriminant of this polynomial and therefore escape the space of hyperlogarithms with rational arguments; $P_{7,11}$ is not linearly reducible as such. But \eqref{eq:P711-d10} can be linearized: If we change variables according to
$\SP_3' \SP_1 = \SP_3 (\SP_1 + \SP_2 + \SP_4)$, $\SP_4 = \SP_4'(\SP_2 + \SP_3')$ and $\SP_1 = \SP_1' \SP_4'$, the new denominator
\begin{equation}
	d_{10}'
	= (\SP_2 + \SP_3') (\SP_2 + \SP_2 \SP_4' - \SP_1') (\SP_1' \SP_4' + \SP_2 + \SP_2 \SP_4' + \SP_3' \SP_4')
	\label{eq:P711-d10'}%
\end{equation}
factorizes linearly such that $\SP_1'$, $\SP_3'$ and $\SP_4'$ can indeed be integrated without further complications ($\SP_2 = 1$). Interestingly, the final period is not a multiple zeta value but a linear combination of multiple polylogarithms \eqref{eq:def-Li} evaluated at a sixth roots of unity.

\subsection{Missing alternating sums}
The other two of the most complicated vacuum graphs with $7$ loops (shown in figure~\ref{fig:7loop-primitives}) are linearly reducible without further ado and we succeeded to compute
\begin{align}
\hide{	P_{7,8}
	&= \frac{22383}{20} \mzv{11}
		+ \frac{4572}{5} \left( 
				\mzv{3,5,3}
				- \mzv{3}\mzv{3,5}
			\right)
			-700 \mzv[2]{3} \mzv{5}
			+ 1792 \mzv{3}\left( 
				\frac{27}{80}\mzv{3,5}
  	   + \frac{45}{64} \mzv{3} \mzv{5}
	     - \frac{261}{320} \mzv{8}
			\right)
	\\
}%
	P_{7,9}
	&= \tfrac{92943}{160} \mzv{11}
     + \tfrac{3381}{20} \left(
		 		\mzv{3,5,3}
  	   - \mzv{3}\mzv{3,5} 
			 \right)
     - \tfrac{1155}{4} \mzv[2]{3} \mzv{5}
     + 896 \mzv{3}
    \left(
	    \tfrac{27}{80} \mzv{3,5}
     + \tfrac{45}{64} \mzv{3} \mzv{5}
     - \tfrac{261}{320} \mzv{8}
    \right)
		\label{eq:P79}%
\end{align}
using hyperlogarithms. This result was recently proposed by David Broadhurst \cite{Broadhurst:Radcor2013} and is now confirmed.
Interestingly, for both $P_{7,8}$ and $P_{7,9}$ the last integrand $I_{12}$ has denominator $(\SP_1 + \SP_2)(\SP_1-\SP_2)$ and produces a result involving alternating sums. Only after reducing these to the data mine basis \cite{BluemleinBroadhurstVermaseren:Datamine} we arrived at the multiple zeta value \eqref{eq:P79}. So far, the reason for this absence of alternating sums in massless $\phi^4$-theory still eludes us.

\section{Non-trivial kinematics}
With non-trivial kinematics, the second Symanzik $\phipol$ makes linear reducibility a more restrictive criterion that depends sensitively on the distribution of external momenta and internal masses. In particular it requires at least one propagator to be massless.\footnote{%
This restriction does not apply when there is no momentum dependence at all. Important examples of such graphs that are linearly reducible are known and hyperlogarithms are used to compute generating functions for Mellin moments of massive operator matrix elements \cite{Wissbrock:Massive3loopLadder,AblingerBluemleinRaabSchneiderWissbrock:Hyperlogarithms}.}
\begin{figure}\centering
		$\Graph[0.4]{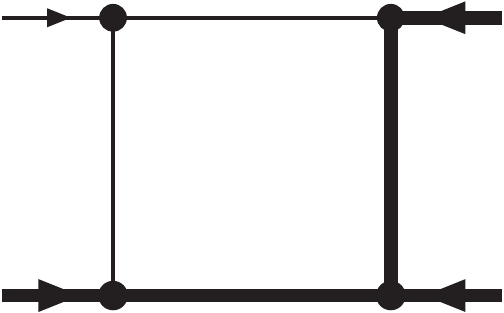}$
		\quad
		$\Graph[0.4]{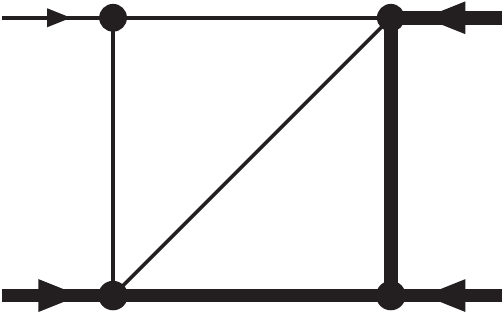}$
		\quad
		$\Graph[0.4]{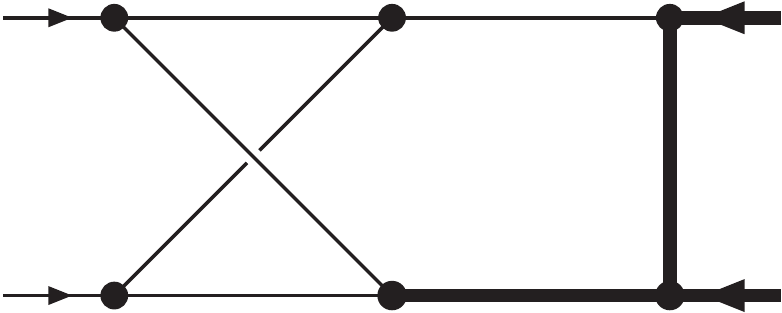}$
		\quad
		$\Graph[0.4]{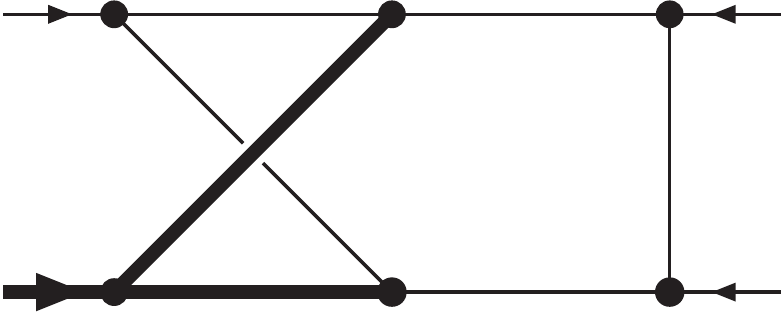}$
		\\
		$\Graph[0.3]{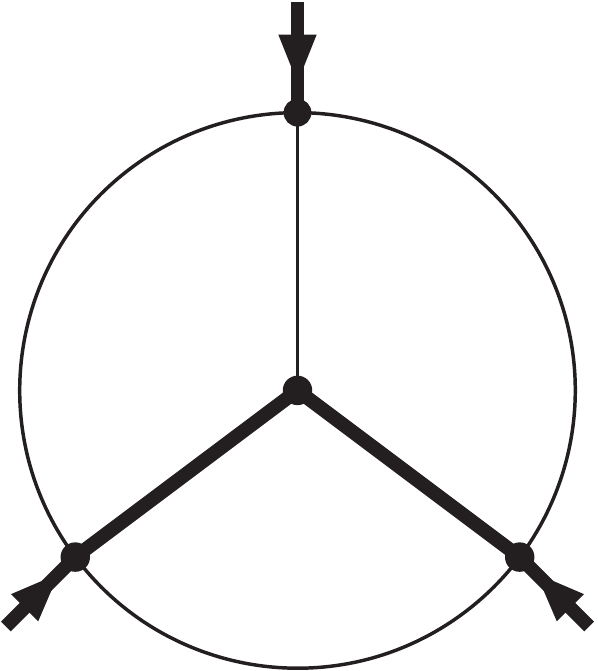}$
		\quad
		$\Graph[0.3]{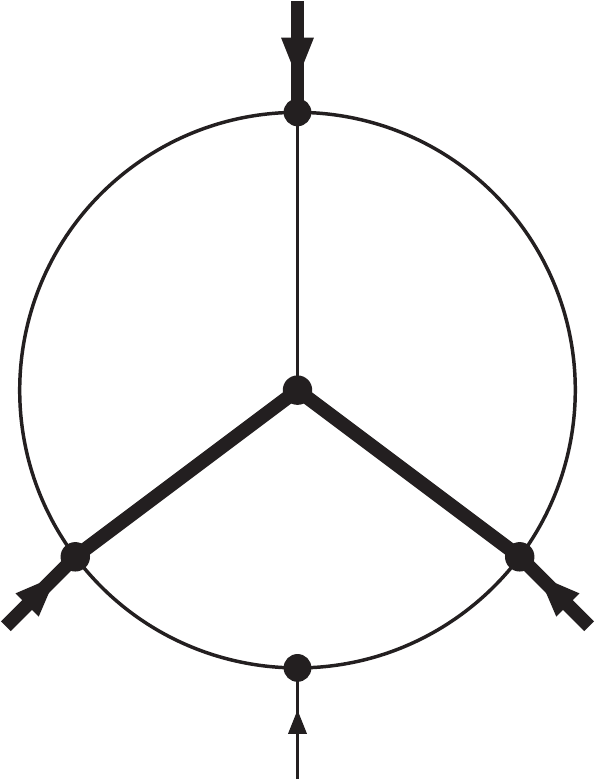}$
		\quad
		$\Graph[0.33]{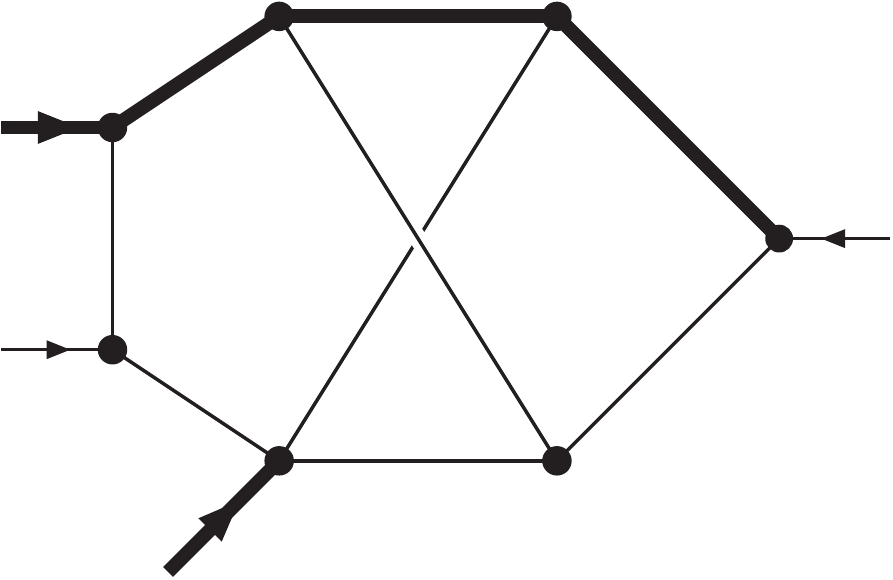}$
		\quad
		$\Graph[0.3]{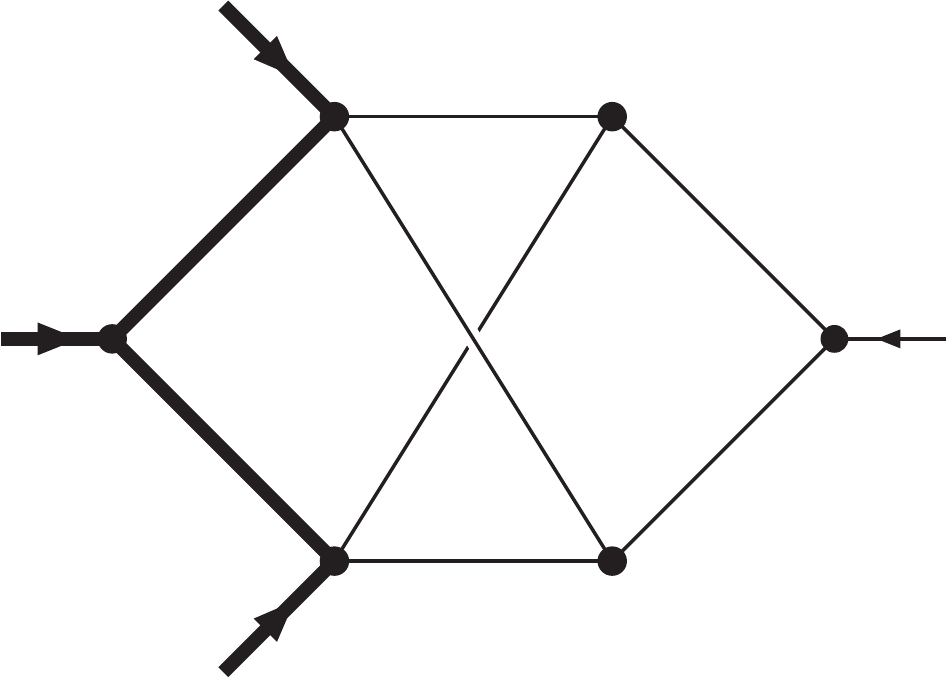}$
	\caption{Examples of linearly reducible graphs with some massive internal and off-shell external momenta (thick edges).}%
	\label{fig:massive}%
\end{figure}

Some examples of reducible graphs with masses are shown in figure~\ref{fig:massive} and some explicit results were computed \cite{Panzer:DivergencesManyScales}. However, no combinatorial characterization of this class of graphs is known so far. We now specialize to two particular kinematic configurations in order to state more general results.
	
\subsection{Massless three-point functions}
Massless three-point functions (depending on arbitrary $p_1^2$, $p_2^2$ and $p_3^2$) seem to be very well suited for parametric integration:
Linear reducibility holds for all such graphs at $2$ loops \cite{ChavezDuhr:Triangles} and even at $3$ loops \cite{Panzer:DivergencesManyScales}. This includes all graphs obtained by removing one three-valent vertex from any of the vacuum graphs in figure~\ref{fig:5loop-vacuum} (external legs are attached to the neighbours of that deleted vertex).

When these integrals are written as hyperlogarithms, one encounters the square root of the K\"{a}ll\'{e}n function $\lambda = \big(p_1^2 + p_2^2 - p_3^2\big)^2 - 4 p_1^2 p_2^2$. The reparametrization given by
\begin{equation}
	p_2^2
	= p_1^2 \cdot z \bar{z}
	\quad\text{and}\quad
	p_3^2
	= p_1^2 \cdot (1-z)(1-\bar{z})
	\quad\text{rationalizes}\quad
	\sqrt{\lambda} = \pm p_1^2 \cdot (z - \bar{z})
	\label{eq:kallen-parameters}%
\end{equation}
in terms of two new variables $z$ and $\bar{z}$. Then $\Phi(G)$ is given by $p_1^{-2\sdd}$ and a rational linear combination of hyperlogarithms $L_{v}(z) L_{w}(\bar{z})$. The polynomial reduction provides an upper bound on the set of letters that may appear in the words $v$ and $w$; put differently it restricts the entries of the \emph{symbol} \cite{GoncharovSpradlinVerguVolovich:ClassicalPolylogarithmsAmplitudesWilsonLoops} of the hyperlogarithms.

Details and explicit results for some integrals can be found in \cite{Panzer:DivergencesManyScales}. There are reducible examples at higher loop orders, but non-reducible graphs appear already at $4$ loops.

\subsection{Massless four-point functions}
\begin{figure}\centering
		$\Graph[0.33]{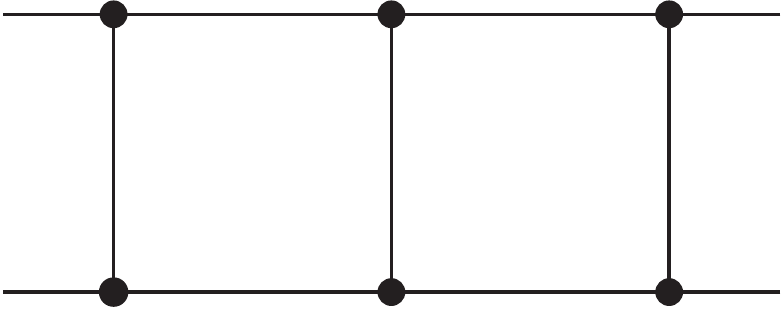}$\quad
		$\Graph[0.33]{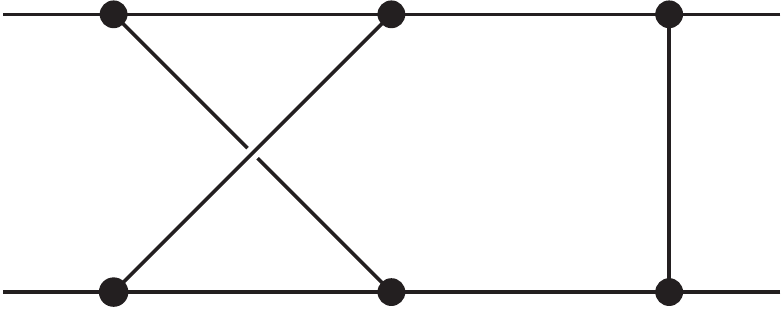}$\quad
		$\Graph[0.33]{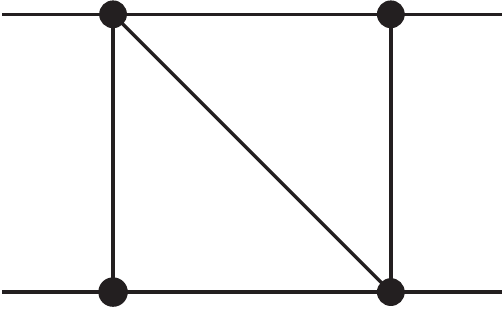}$\quad
		$\Graph[0.33]{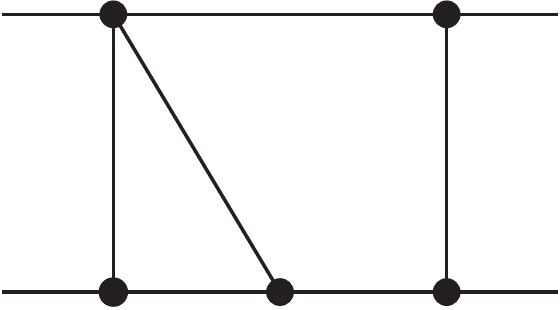}$\quad
		$\Graph[0.33]{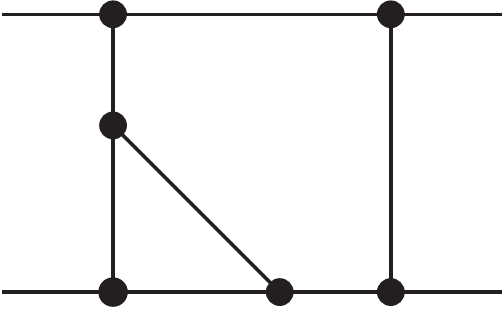}$\quad
		$\Graph[0.33]{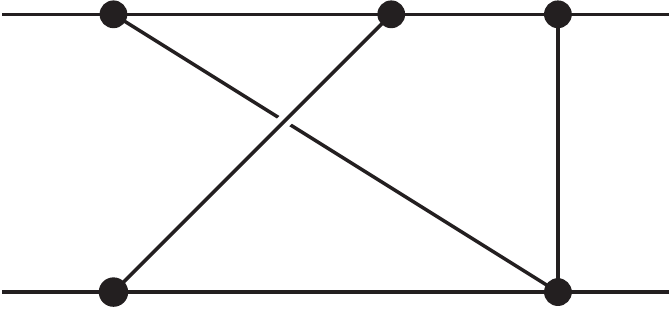}$
	\caption{Two-loop four-point graphs without self-energy (propagator) subgraphs.}%
	\label{fig:2loop-4pt}%
\end{figure}
All massless four-point on-shell graphs ($p_1^2=p_2^2=p_3^2=p_4^2=0$) with at most two loops are linearly reducible \cite{BognerLueders:MasslessOnShell}. In particular these include the graphs of figure~\ref{fig:2loop-4pt}.

At $3$ loops the first graphs that are not linearly reducible occur, including the complete graph $K_4$ (first graph in bottom row of figure~\ref{fig:massive} with the fourth external momentum attached to the center) which was recently evaluated \cite{HennSmirnov:K4} to harmonic polylogarithms \cite{RemiddiVermaseren:HarmonicPolylogarithms}. For its parametric integration, a change of variables is necessary \cite{Panzer:DivergencesManyScales}.

\begin{figure}\centering
	$B_3 = \Graph[0.4]{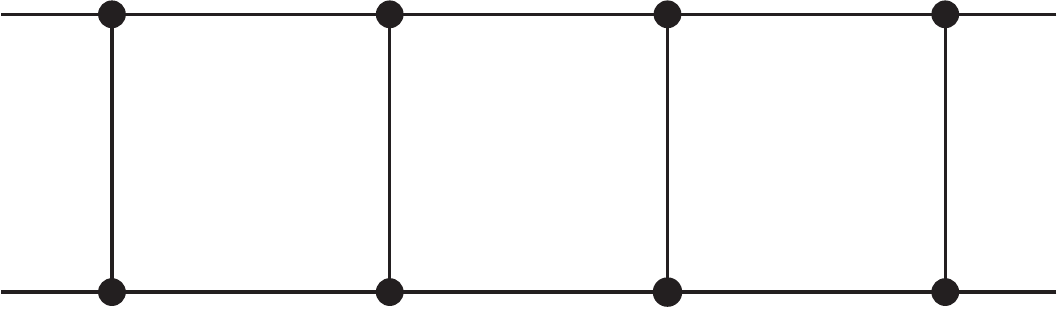}$
	\qquad
	$B_4 = \Graph[0.4]{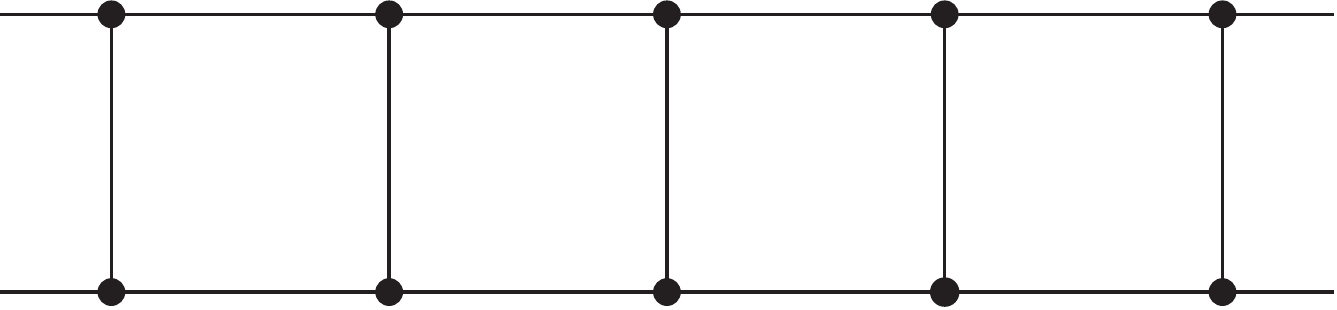}$
	\caption{Examples of the \emph{box ladder} graphs $B_n$ of $n=3$ and $n=4$ loops.} %
	\label{fig:ladderboxes} %
\end{figure}

While the general type of massless four-point functions is not clear, it seems that at least all of the $n$-loop box ladders $B_n$ (figure~\ref{fig:ladderboxes}) are linearly reducible: For on-shell kinematics we always obtained a result in terms of harmonic polylogarithms. Due to the recent interest \cite{Kazakov:MultiBoxSix} into these integrals in $\dimension=6$ dimensions (where they are finite), we list the values $c_n \defas \restrict{\Phi(B_n)}{s=1, t=0}$ for $n \leq 6$ where $s = (p_1+p_2)^2$ and $t=(p_1+p_4)^2$ measure the momentum running through $B_n$ in the horizontal and vertical directions:\footnote{Printing the expressions for $\Phi(B_n)$ including the dependence on $s$ and $t$ would take too much space.}
\begin{align}
c_2 &= 2\mzv{2}
	,\label{eq:box-2} %
\\
c_3 &= 4 \mzv[2]{3} 
	+ \tfrac{124}{35}\mzv[3]{2}
	- 8 \mzv{3}
	- 6\mzv{2}
	,\label{eq:box-3} %
\\
c_4 &=
 - 56\mzv{7}
 - 32\mzv{2}\mzv{5}
 + 32\mzv[2]{3}
 + \tfrac{8}{5}\mzv{3}
\left(
4\mzv[2]{2}
 - 15
\right)
 + \tfrac{992}{35}\mzv[3]{2}
 - 8\mzv[2]{2}
 - 18\mzv{2}
 ,\label{eq:box-4} %
\\
c_5 &=
	56\mzv{7}
	\left(
 \mzv{3} 
 - 5
\right)
 + 26\mzv[2]{5}
 + 4\mzv{5}
\left(
 8\mzv{2}\mzv{3}
 + 35\mzv{3}
 - 40\mzv{2}
 - 49
\right)
 + \tfrac{4}{5}\mzv[2]{3}
\left(
 140
 - 25\mzv{2}
 - 4\mzv[2]{2}
\right)
\nonumber\\&\quad
 + 8\mzv{3}
\left(
7\mzv{2}
 + 4\mzv[2]{2}
 - 14
\right)
 - \tfrac{1168}{385}\mzv[5]{2}
 - \tfrac{24}{7}\mzv[4]{2}
 + \tfrac{496}{5}\mzv[3]{2}
 + 4\mzv{2}
\left(
 2\mzv{3,5}
 - 21
\right)
 + 20\mzv{3,5}
 + 4\mzv{3,7}
 ,\label{eq:box-5} %
\\
 c_6 &=
 \tfrac{18864}{35}\mzv[3]{2}
 + 336\mzv{3,5}
 - 12\mzv{9}
\left(
20\mzv{2}
 + 161
\right)
 + \tfrac{8}{5}\mzv{7}
\left(
104\mzv[2]{2}
 + 35\mzv{2}
 + 840\mzv{3}
 - 1120
\right)
\nonumber\\&\quad
 + 624\mzv[2]{5}
 + \tfrac{16}{35}\mzv{5}
\left(
1680\mzv{2}\mzv{3}
 - 3675
 - 12\mzv[3]{2}
 - 2240\mzv{2}
 + 490\mzv[2]{2}
 + 5145\mzv{3}
\right)
\nonumber\\&\quad
 + 96 \left( \mzv[2]{2} + \mzv{3,7} \right)
 - \tfrac{48}{5}\mzv[2]{3}
\left(
35\mzv{2}
 + 8\mzv[2]{2}
 - 60
\right)
 - \tfrac{32}{5}\mzv{3}
\left(
105
 - 32\mzv[2]{2}
 + 3\mzv[3]{2}
 - 75\mzv{2}
\right)
\nonumber\\&\quad
  +24\mzv{2}
	\left(
		 8\mzv{3,5}
		 - 21
	\right)
   - \tfrac{28032}{385}\mzv[5]{2}
   - \tfrac{288}{5}\mzv[4]{2}
  - 1320\mzv{11}
  .\label{eq:box-6} %
\end{align}

\section{Problems for parametric integration with hyperlogarithms}
\begin{enumerate}
	\item	
To compute divergent integrals in dimensional regularization, one first needs to construct a representation involving only convergent integrands. An algorithm that solves this problem was presented in \cite{Panzer:DivergencesManyScales}, but already for low numbers of divergences it can produce expressions that become untractibly large. It seems very promising to combine this algorithm with programs for integration by parts, in order to obtain a reduction to \emph{finite (convergent) master integrals}.

	\item
We know many cases (like $K_4$ and $P_{7,11}$ mentioned above) of integrals that are not linearly reducible in the original Schwinger parameters, but become so after a suitable change of variables. It is unclear under which general circumstances this is possible. So far, there is only one combinatorial analysis available which considers a particular kind of transformation for vacuum integrals \cite{Yeats:SomeInterpretations}.
	
	\item 
Why do alternating sums so far not occur in massless propagators?

	\item
The implementation \cite{Panzer:HyperIntAlgorithms} works in the Euclidean region. Analytic continuation to various kinematic regimes of the physical region can in general be very cumbersome and should be automated.
\end{enumerate}

\bibliographystyle{JHEPsortdoi}
\bibliography{../../qft}

\providecommand{\href}[2]{#2}\begingroup\raggedright\begin{thebibliography}{10}

\bibitem{Wissbrock:Massive3loopLadder}
J.~Ablinger, J.~Bl\"{u}mlein, A.~Hasselhuhn, S.~Klein, C.~Schneider, and
  F.~Wi{\ss}brock, {\it Massive 3-loop ladder diagrams for quarkonic local
  operator matrix elements},
  \href{http://dx.doi.org/\detokenize{10.1016/j.nuclphysb.2012.06.007}}{{\em
  Nucl. Phys. B} {\bf 864} (Nov., 2012) 52--84},
  [\href{http://xxx.lanl.gov/abs/1206.2252}{{\tt arXiv:1206.2252}}].

\bibitem{AblingerBluemleinRaabSchneiderWissbrock:Hyperlogarithms}
J.~{Ablinger}, J.~{Bl{\"u}mlein}, C.~{Raab}, C.~{Schneider}, and
  F.~{Wi{\ss}brock}, {\it Calculating massive 3-loop graphs for operator matrix
  elements by the method of hyperlogarithms},
  \href{http://dx.doi.org/\detokenize{10.1016/j.nuclphysb.2014.04.007}}{{\em
  Nucl. Phys. B} {\bf 885} (Aug., 2014) 409--447},
  [\href{http://xxx.lanl.gov/abs/1403.1137}{{\tt arXiv:1403.1137}}].

\bibitem{AdamsBognerWeinzierl:Sunrise}
L.~Adams, C.~Bogner, and S.~Weinzierl, {\it The two-loop sunrise graph with
  arbitrary masses},
  \href{http://dx.doi.org/\detokenize{10.1063/1.4804996}}{{\em Journal of
  Mathematical Physics} {\bf 54} (May, 2013) 052303},
  [\href{http://xxx.lanl.gov/abs/1302.7004}{{\tt arXiv:1302.7004}}].

\bibitem{BaikovChetyrkin:FourLoopPropagatorsAlgebraic}
P.~A. Baikov and K.~G. Chetyrkin, {\it Four loop massless propagators: {An}
  algebraic evaluation of all master integrals},
  \href{http://dx.doi.org/\detokenize{10.1016/j.nuclphysb.2010.05.004}}{{\em
  Nucl. Phys. B} {\bf 837} (Oct., 2010) 186--220},
  [\href{http://xxx.lanl.gov/abs/1004.1153}{{\tt arXiv:1004.1153}}].

\bibitem{BlochVanhove:Sunset}
S.~Bloch and P.~Vanhove, {\it The elliptic dilogarithm for the sunset graph},
  \href{http://xxx.lanl.gov/abs/1309.5865}{{\tt arXiv:1309.5865}}.

\bibitem{BluemleinBroadhurstVermaseren:Datamine}
J.~{Bl{\"u}mlein}, D.~J. {Broadhurst}, and J.~A.~M. {Vermaseren}, {\it {The
  Multiple Zeta Value data mine}},
  \href{http://dx.doi.org/\detokenize{10.1016/j.cpc.2009.11.007}}{{\em Comput.
  Phys. Commun.} {\bf 181} (Mar., 2010) 582--625},
  [\href{http://xxx.lanl.gov/abs/0907.2557}{{\tt arXiv:0907.2557}}].

\bibitem{BognerBrown:SymbolicIntegration}
C.~Bogner and F.~C.~S. Brown, {\it Symbolic integration and multiple
  polylogarithms},  \pos{PoS(LL20129)053}
  [\href{http://xxx.lanl.gov/abs/1209.6524}{{\tt arXiv:1209.6524}}].

\bibitem{BognerLueders:MasslessOnShell}
C.~Bogner and M.~{L\"{u}ders}, {\it Multiple polylogarithms and linearly
  reducible {Feynman} graphs},  \href{http://xxx.lanl.gov/abs/1302.6215}{{\tt
  arXiv:1302.6215}}.

\bibitem{BognerWeinzierl:GraphPolynomials}
C.~Bogner and S.~Weinzierl, {\it {Feynman Graph Polynomials}},
  \href{http://dx.doi.org/\detokenize{10.1142/S0217751X10049438}}{{\em
  International Journal of Modern Physics A} {\bf 25} (2010) 2585--2618},
  [\href{http://xxx.lanl.gov/abs/1002.3458}{{\tt arXiv:1002.3458}}].

\bibitem{Broadhurst:Radcor2013}
D.~J. Broadhurst, {\it The number theory of radiative corrections}, \href{https://conference.ippp.dur.ac.uk/getFile.py/access?contribId=36&sessionId=12&resId=0&materialId=slides&confId=341}{in {\em
  RADCOR 2013}, (Lumley Castle, UK), Sept., 2013}.

\bibitem{BroadhurstKreimer:KnotsNumbers}
D.~J. Broadhurst and D.~Kreimer, {\it Knots and numbers in $\phi^4$ theory to 7
  loops and beyond},
  \href{http://dx.doi.org/\detokenize{10.1142/S012918319500037X}}{{\em Int. J.
  Mod. Phys. C} {\bf 6} (Aug., 1995) 519--524},
  [\href{http://xxx.lanl.gov/abs/hep-ph/9504352}{{\tt hep-ph/9504352}}].

\bibitem{Brown:PeriodsFeynmanIntegrals}
F.~C.~S. Brown, {\it On the periods of some {Feynman} integrals},
  \href{http://xxx.lanl.gov/abs/0910.0114}{{\tt arXiv:0910.0114}}.

\bibitem{Brown:TwoPoint}
F.~C.~S. Brown, {\it {The Massless Higher-Loop Two-Point Function}},
  \href{http://dx.doi.org/\detokenize{10.1007/s00220-009-0740-5}}{{\em Commun.
  Math. Phys.} {\bf 287} (May, 2009) 925--958},
  [\href{http://xxx.lanl.gov/abs/0804.1660}{{\tt arXiv:0804.1660}}].

\bibitem{BrownSchnetz:K3phi4}
F.~C.~S. Brown and O.~Schnetz, {\it A {K3} in $\phi^{4}$},
  \href{http://dx.doi.org/\detokenize{10.1215/00127094-1644201}}{{\em Duke
  Math. J.} {\bf 161} (July, 2012) 1817--1862},
  [\href{http://xxx.lanl.gov/abs/1006.4064}{{\tt arXiv:1006.4064}}].

\bibitem{BrownSchnetz:ZigZag}
F.~C.~S. Brown and O.~Schnetz, {\it Proof of the zig-zag conjecture},
  \href{http://xxx.lanl.gov/abs/1208.1890}{{\tt arXiv:1208.1890}}.

\bibitem{BrownYeats:SpanningForestPolynomials}
F.~C.~S. Brown and K.~A. Yeats, {\it {Spanning Forest Polynomials and the
  Transcendental Weight of Feynman Graphs}},
  \href{http://dx.doi.org/\detokenize{10.1007/s00220-010-1145-1}}{{\em Commun.
  Math. Phys.} {\bf 301} (Jan., 2011) 357--382},
  [\href{http://xxx.lanl.gov/abs/0910.5429}{{\tt arXiv:0910.5429}}].

\bibitem{CaronHuotLarsen:UniquenessTwoLoopMasterContours}
S.~{Caron-Huot} and K.~J. Larsen, {\it Uniqueness of two-loop master contours},
   \href{http://dx.doi.org/\detokenize{10.1007/JHEP10(2012)026}}{{\em JHEP}
  {\bf 2012} (Oct., 2012) 26}, [\href{http://xxx.lanl.gov/abs/1205.0801}{{\tt
  arXiv:1205.0801}}].

\bibitem{ChavezDuhr:Triangles}
F.~Chavez and C.~Duhr, {\it Three-mass triangle integrals and single-valued
  polylogarithms},
  \href{http://dx.doi.org/\detokenize{10.1007/JHEP11(2012)114}}{{\em JHEP} {\bf
  11} (Nov., 2012) 114}, [\href{http://xxx.lanl.gov/abs/1209.2722}{{\tt
  arXiv:1209.2722}}].

\bibitem{Chen:IteratedPathIntegrals}
K.~T. Chen, {\it Iterated path integrals},
  \href{http://dx.doi.org/\detokenize{10.1090/S0002-9904-1977-14320-6}}{{\em
  Bull. Amer. Math. Soc.} {\bf 83} (Sept., 1977) 831--879}.

\bibitem{GoncharovSpradlinVerguVolovich:ClassicalPolylogarithmsAmplitudesWilsonLoops}
A.~B. Goncharov, M.~Spradlin, C.~Vergu, and A.~Volovich, {\it {Classical
  Polylogarithms for Amplitudes and Wilson Loops}},
  \href{http://dx.doi.org/\detokenize{10.1103/PhysRevLett.105.151605}}{{\em
  Phys. Rev. Lett.} {\bf 105} (Oct., 2010) 151605},
  [\href{http://xxx.lanl.gov/abs/1006.5703}{{\tt arXiv:1006.5703}}].

\bibitem{HennSmirnov:K4}
J.~M. Henn, A.~V. Smirnov, and V.~A. Smirnov, {\it Evaluating single-scale
  and/or non-planar diagrams by differential equations},
  \href{http://dx.doi.org/\detokenize{10.1007/JHEP03(2014)088}}{{\em JHEP} {\bf
  2014} (Mar., 2014) } TTP13-046,
  [\href{http://xxx.lanl.gov/abs/1312.2588}{{\tt arXiv:1312.2588}}].

\bibitem{Kazakov:MultiBoxSix}
D.~I. Kazakov, {\it {Evaluation of Multi-Box Diagrams in Six Dimensions}},
  \href{http://dx.doi.org/\detokenize{10.1007/JHEP04(2014)121}}{{\em JHEP} {\bf
  2014} (Apr., 2014) 1--9}, [\href{http://xxx.lanl.gov/abs/1402.1024}{{\tt
  arXiv:1402.1024}}].

\bibitem{LappoDanilevsky}
J.~A. Lappo-Danilevsky, {\em M\'{e}moires sur la th\'{e}orie des syst\`{e}mes
  des \'{e}quations diff\'{e}rentielles lin\'{e}aires}, vol.~I--III.
\newblock Chelsea, 1953.

\bibitem{Maple}
{Maplesoft, a division of Waterloo Maple Inc.}, ``Maple 16.''

\bibitem{Nakanishi:GraphTheoryFeynmanIntegrals}
N.~Nakanishi, {\em {Graph} theory and {Feynman} integrals}, vol.~11 of {\em
  Mathematics and its applications}.
\newblock Gordon and Breach, New York, 1971.

\bibitem{NandanPaulosSpradlinVolovich:StarIntegrals}
D.~Nandan, M.~F. Paulos, M.~Spradlin, and A.~Volovich, {\it Star integrals,
  convolutions and simplices},
  \href{http://dx.doi.org/\detokenize{10.1007/JHEP05(2013)105}}{{\em JHEP} {\bf
  2013} (May, 2013) 105}, [\href{http://xxx.lanl.gov/abs/1301.2500}{{\tt
  arXiv:1301.2500}}].

\bibitem{Panzer:MasslessPropagators}
E.~Panzer, {\it On the analytic computation of massless propagators in
  dimensional regularization},
  \href{http://dx.doi.org/\detokenize{10.1016/j.nuclphysb.2013.05.025}}{{\em
  Nucl. Phys. B} {\bf 874} (Sept., 2013) 567--593},
  [\href{http://xxx.lanl.gov/abs/1305.2161}{{\tt arXiv:1305.2161}}].

\bibitem{Panzer:HyperIntAlgorithms}
E.~Panzer, {\it Algorithms for the symbolic integration of hyperlogarithms with
  applications to {Feynman} integrals},
  \href{http://xxx.lanl.gov/abs/1403.3385}{{\tt arXiv:1403.3385}}.

\bibitem{Panzer:DivergencesManyScales}
E.~Panzer, {\it On hyperlogarithms and {F}eynman integrals with divergences and
  many scales},
  \href{http://dx.doi.org/\detokenize{10.1007/JHEP03(2014)071}}{{\em JHEP} {\bf
  2014} (Mar., 2014) 71}, [\href{http://xxx.lanl.gov/abs/1401.4361}{{\tt
  arXiv:1401.4361}}].

\bibitem{RemiddiVermaseren:HarmonicPolylogarithms}
E.~Remiddi and J.~A.~M. Vermaseren, {\it Harmonic polylogarithms},
  \href{http://dx.doi.org/\detokenize{10.1142/S0217751X00000367}}{{\em Int. J.
  Mod. Phys. A} {\bf 15} (2000), no. 5 725--754},
  [\href{http://xxx.lanl.gov/abs/hep-ph/9905237}{{\tt hep-ph/9905237}}].

\bibitem{Schnetz:Census}
O.~Schnetz, {\it Quantum periods: {A} {Census} of $\phi^4$-transcendentals},
  \href{http://dx.doi.org/\detokenize{10.4310/CNTP.2010.v4.n1.a1}}{{\em Commun.
  Number Theory Phys.} {\bf 4} (2010), no. 1 1--47},
  [\href{http://xxx.lanl.gov/abs/0801.2856}{{\tt arXiv:0801.2856}}].

\bibitem{Schnetz:GraphicalFunctions}
O.~Schnetz, {\it Graphical functions and single-valued multiple
  polylogarithms},  \href{http://xxx.lanl.gov/abs/1302.6445}{{\tt
  arXiv:1302.6445}}.

\bibitem{SmirnovTentyukov:FIESTA}
A.~V. Smirnov and M.~N. Tentyukov, {\it {Feynman Integral Evaluation by a
  Sector decomposiTion Approach (FIESTA)}},
  \href{http://dx.doi.org/\detokenize{10.1016/j.cpc.2008.11.006}}{{\em Comput.
  Phys. Commun.} {\bf 180} (2009) 735--746},
  [\href{http://xxx.lanl.gov/abs/0807.4129}{{\tt arXiv:0807.4129}}].

\bibitem{Yeats:SomeInterpretations}
K.~A. Yeats, {\it Some combinatorial interpretations in perturbative quantum
  field theory},  \href{http://xxx.lanl.gov/abs/1302.0080}{{\tt
  arXiv:1302.0080}}.

\end{thebibliography}\endgroup

\end{document}